\documentclass[twocolumn,aps,amsmath,amssymb,]{revtex4-2} 
\usepackage{graphicx}
\usepackage{dcolumn}
\usepackage{bm}
\usepackage[T1]{fontenc}
\usepackage{xcolor}
\usepackage{amsthm}

\begin{document}

\title{Two-qubit quantum gates with minimal pulse sequences}

\author{Ignacio R. Sola}
\email{corresponding author: isolarei@ucm.es}
 \affiliation{Departamento de Quimica Fisica, Universidad Complutense, 28040 Madrid, Spain}
\author{Seokmin Shin}
 \affiliation{School of Chemistry, Seoul National University, 08826 Seoul, Republic of Korea}
\author{Bo Y. Chang}
 \email{corresponding author: boyoung@snu.ac.kr}
 \affiliation{School of Chemistry, Seoul National University, 08826 Seoul, Republic of Korea}
 \affiliation{Departamento de Quimica Fisica, Universidad Complutense, 28040 Madrid, Spain}

\begin{abstract}
 Working with trapped atoms at close distance to each other, we show that one can implement entangling gates based on non-independent qubits using a single pulse per qubit, or a single structured pulse. The optimal parameters depend on approximate solutions of Diophantine equations, causing the fidelity to never be exactly perfect, even under ideal conditions, although the errors can be made arbitrarily smaller at the cost of stronger fields. We fully characterize the mechanism by which the gates operate, and show that the main source of error in realistic implementations comes from fluctuations in the peak intensity, which especially damages the fidelity of the gates that use stronger fields. Working with two-pulse sequences, instead of one, enables the use of a plethora of mechanisms and a broad range of optimal parameters to choose from, to achieve high-fidelity gates.
\end{abstract}

\maketitle

\section{Introduction}

Most quantum control protocols rely on complex pulse sequences or
pulse structures in the time domain. We show in this work that,
for ordered systems with a high degree of control in their spatial structure, 
it is possible to use the simplest pulse sequences and
achieve the same level of control acting on the spatial 
degrees of freedom, adding some complexity 
in the spatial domain. 

Quantum computers are the paramount systems where one needs a 
maximum degree of control over their spatial and time domain
properties to minimize the effects of decoherence, and to 
synchronize the different interference effects that are involved
in the speed-up properties of quantum algorithms~\cite{Steuerman_Science2021, Lieven_Nature2022, Wu_PRL2021, Wright_NatureComm2019, Borzenkova_APL2021, Saffman_Nature2022, Cirac_Nature2000, Ladd_Nature2010, Schoelkopf_Science2013, Kelly_Science2015, Harty_PRL2014, Wrachtrup_PRL2004,Saffman_RMP2010}.
Atoms trapped by optical tweezers~\cite{Browaeys_PRX2014, Browaeys_Science2016, Wilson_PRL2022, Thompson_PRXQ2022, Ahn_OptE2016}, using highly excited Rydberg states for dipole-blockaded interactions~\cite{Comparat_JOSAB2010, Tong_PRL2004, Saffman_NPhys2009, Grangier_Nphys2009, Adams_PRL2010}, are one of the promising platforms for quantum computing, due to their extended coherence times~\cite{Saffman_RMP2010}, strong and long-range interactions~\cite{Saffman_RMP2010}, scalability~\cite{Browaeys_PRX2014, Browaeys_Nature2018}, and addressability~\cite{Kuzmich_PRL2022, Kuzmich_Science2012, Adams_JPB2019, Adams_PRL2010, Ohmori_NP2022}.
This adaptability makes Rydberg atoms a versatile resource for implementing 
multi-particle entanglement~\cite{Lukin_PRL2018, Zhan_PRL2017, Ahn_PRL2020, Grangier_PRL2010, Saffman_Nature2022, Saffman_PRA2015, Saffman_PRA2010, Picken_QCT2018, Malinovsky_PRA2004, Malinovsky_PRL2004, Malinovsky_PRL2006}, simple quantum circuits~\cite{Jaksch_PRL2000, Saffman_QIP2011, Saffman_PRA2015, Lukin_PRL2001, Lukin_PRL2019, Cohen_PRXQ2021, Shi_QSTech2022, Shi_PRApp2018, Adams_PRL2014, Adams_JPB2019, Malinovsky_PRA2014, Goerz_JPB2011, Morgado_AVSQSci2021, Alexey_PRL2021, Sanders_PRA2020} and even
quantum gates across different quantum computing platforms~\cite{Hennrich_Nature2020, Molmer_PRX2020, Khazali_OE2023, Pohl_Quantum2022, Raithel_PRL2011}.

Current technology allows to control the position and spatial organization 
of the atoms in atomic traps with great precision, and this property has been
extensively used for quantum simulations and to prepare different entangled states~\cite{Ahn_PRR2021, Ahn_PRXQ2020}. 
Most quantum circuits, however, have relied on the use of independent qubits,
which for homogeneous qubits impose large interatomic distances and hence operate
with weak dipole blockades, leading to slow two-qubit gates.
Several C-PHASE~\cite{Jaksch_PRL2000,Goerz_JPB2011,Goerz_PRA2014} and C-NOT~\cite{Saffman_QIP2011} 
gate proposals reported implementation times in the microsecond.

Since the ancillary states are highly excited (although long-lived) Rydberg states,
speeding-up the processes has obvious advantages, as it drastically reduces the effect
of decoherence.
But this typically requires working with closer, and hence non-independent, qubits,
which brings an additional level of control in the atomic positions and the spatial
profiles of the laser beams, for which we proposed a novel spatio-temporal control
framework~\cite{Sola_Nanoscale2023, previouswork, Sola_unpublished_JCP}.
It turns out that by addressing both qubits at the same time using structured light 
and controlling the amplitude of the fields at the location of each qubit, one can extend
the well-known scheme proposed by Jacksch et al.~\cite{Jaksch_PRL2000} with minimal
changes, but working in the nanosecond regime, at least under ideal conditions~\cite{Sola_Nanoscale2023}.
The scheme, called the SOP (symmetrically orthogonal protocol) prepared a coherent dark
state to transition the population through Rydberg states, isolating the effects of odd and even
pulses in the pulse sequence, which added to the effect of the dipole blockade~\cite{Sola_Nanoscale2023}.
But by breaking the symmetry of the system with apparent disorder and fully controlling the 
spatial profile of the lasers, we showed that a multitude of schemes could implement the
CZ gate with higher fidelity, in 2-qubit~\cite{previouswork} and N-qubit systems~\cite{Sola_unpublished_JCP}.

Alternatively, there have been recent promising results addressing
two or three qubits in symmetric arrangements of the atoms, which
correspond to a very specific scenario from our setup of possible
arrangements. Here, the control is enhanced by phase modulation of the pulses~\cite{Sanders_PRA2020,Jandura_Quantum2022}, so all the pulse
complexity lies again in the time-domain.  

It is possible to classify the optimal control protocols obtained by numerical algorithms
and to analyze the correlations among subsets of control parameters. In particular,
we found highly constraint optimal parameters in protocols that use two-pulse 
sequences~\cite{previouswork}.
In this work, we focus on the minimal pulse sequences, where all the 
control practically depends only on the spatial domain.
In particular, we find that for non-independent qubits, there are solutions that require a 
single pulse, which depends on approximate solutions of Diophantine equations. 
By scrutinizing the nature of two-pulse sequences, we determine the set of
possible protocols and analyze the working principles behind their dynamics.
In this work, we also propose a different physical realization of the non-independent qubit gates, 
using superposed Gaussian beams, and provide an analysis of the role of the
fluctuation and noise in the different control parameters on the robustness of the protocols.

\section{Setup}

\subsection{Dynamics}

We study here gate protocols based on non-independent qubits,
that operate with pulses
that interact with both qubits (or more than one qubit in
the general setup) at the same time.
Then one must control both the temporal features of the pulse
sequence (pulse areas, frequencies, relative phases) as well as
the spatial properties of the pulse beams.

An example is the SOP scheme~\cite{Sola_Nanoscale2023},
where one applies a sequence of three structured pulses, 
using hybrid modes of light ({\em e.g.} superposition of TEM modes),
with different amplitudes at the qubit sites:
$\Omega_k(\vec{r}_A,t) = a_k\mu_{0r} E_k(t)/\hbar = a_k \Omega_k(t)$,
$\Omega_k(\vec{r}_B,t) = b_k\mu_{0r} E_k(t)/\hbar = a_k \Omega_k(t)$.
The first pulse has a large amplitude
on qubit $A$, $a_1$, and a smaller amplitude on qubit $B$, $b_1$. The second
one reverts the role, but with a phase shift in one amplitude: $a_2 = -b_1$, and $b_2 = a_1$. Finally, the
third pulse is a replica of the first one.
The role of the $a$ and $b$ coefficients can be obviously interchanged.
Arranging the factors that participate on the local amplitudes 
(henceforth called geometrical factors)
as components of vectors ${\bf e}_k$ (henceforth structural vectors), 
then we observe that ${\bf e}_1 {\bf e}_2 = 0$
and  ${\bf e}_1 {\bf e}_3 = 1$.
The geometrical factors can be partially incorporated into the Franck-Condon factors 
$\mu_{0r}$, so one can assume, without loss of generality, that $a_k$ and $b_k$ 
are normalized to unity ($|{\bf e}_k| = \sqrt{a_k^2 + b_k^2} = 1$). 

For atoms a short distance apart, the dipole blockade forbids 
that more than one Rydberg state can be populated during the laser action.
In the simplest model that describes the two-qubit gate \cite{previouswork}, 
the system is described by $8$ states: the computational basis and
ancillary states with Rydberg excitations, as the pulse frequencies
are chosen to be in resonance with the $|0\rangle \rightarrow |r\rangle$
transition\footnote{The scheme performs similarly
using pulses in resonance from the $|1\rangle$ state to the Rydberg state.}. 
The Hamiltonian is block-diagonal for each computational basis
${\sf H}_k^V \oplus {\sf H}_k^A \oplus {\sf H}_k^B \oplus {\sf H}^D$,
where 
$${\sf H}_k^V = -\frac{1}{2} \Omega_k(t) \left(
a_k|00\rangle \langle r0| + b_k |00\rangle
\langle 0r| + \mathrm{h.c.} \right)$$
is the Hamiltonian of a $3$-level subsystem in $V$ configuration,
acting  in the subspace of $\lbrace |00\rangle, |r0\rangle, |0r\rangle \rbrace$
states, 
${\sf H}_k^A = -\frac{1}{2} a_k\Omega_k(t) \left(  
|01\rangle\langle r1| + \mathrm{h.c.} \right)$
and 
${\sf H}_k^B = -\frac{1}{2} b_k\Omega_k(t) \left(  
|10\rangle\langle 1r| + \mathrm{h.c.} \right)$ 
are two-level Hamiltonians acting 
in the subspace of $\lbrace |01\rangle, |r1\rangle \rbrace$ and
$\lbrace |10\rangle, |1r\rangle \rbrace$ respectively.
We will refer generally to any of these subsystems with the superscript $S$
($S = V,  A, B$).
Finally, ${\sf H}^D = 0 |11\rangle\langle 11|$
is the Hamiltonian acting on the double-excited qubit state $|11\rangle$, decoupled from any field.


Using temporally non-overlapping pulses,
the propagator for the time evolution
is the time-ordered product of the evolution operators for each pulse, 
${\sf U}^{S} = \prod_{k=0}^{N_p-1} U^{S}_{N_p-k}$, which is analytical.
For the $V$ subsystem, 
\begin{equation}
U^{V}_{k} = 
\left( \begin{array}{ccc}
\cos \theta^V_k & i  a_{k} \sin \theta^V_k & i b_{k} \sin \theta^V_k  \\
i a_{k} \sin \theta^V_k & a_{k}^2 \cos \theta^V_k + b_k^2 & 
a_{k}b_{k} \left[ \cos \theta^V_k - 1 \right]  \\
i b_{k} \sin \theta^V_k & a_{k}b_{k} \left[ \cos \theta^V_k - 1 \right]  & 
b_{k}^2 \cos \theta^V_k + a_k^2 \end{array} \right)  \label{UV}
\end{equation}
where the mixing angle
$$\theta^V_k = \frac{1}{2} \int_{-\infty}^{\infty}\Omega_{k}(t) dt = \frac{1}{2} A_k$$
is half the pulse area.
For the two-level subsystems $A$ and $B$, we can use the same expression for
the relevant states with
$a_k = 1, b_k = 0$, for $U_k^A$, and vice versa for $U_k^B$.
However, the mixing angles depend on the local coupling:
$\theta^A_k = a_k A_k / 2$ and $\theta_k^B = b_k A_k / 2$.
We will refer to the generalized pulse areas, $2\theta^S_k$, as GPA.

The SOP uses spatially orthogonal vectors such that the state of the system after the
first pulse acting on $|00\rangle$, is a dark state of the Hamiltonian for the second pulse ${\sf H}_2^V$,
so the second pulse does not affect this state. In this way, the SOP works similarly to the
JP, but with non-independent qubits.
In this work, we will study families of schemes that can operate with even fewer pulses, although they
typically require the same (or larger) accumulated pulse area, 
$A_T = \sum_k |A_k|$.
In the following section, we propose a possible scheme to control the structural factors over a wide
range of values (including negative factors) by using superposed laser beams.

\subsection{Implementation}

\begin{figure}
\includegraphics[width=8.5cm]{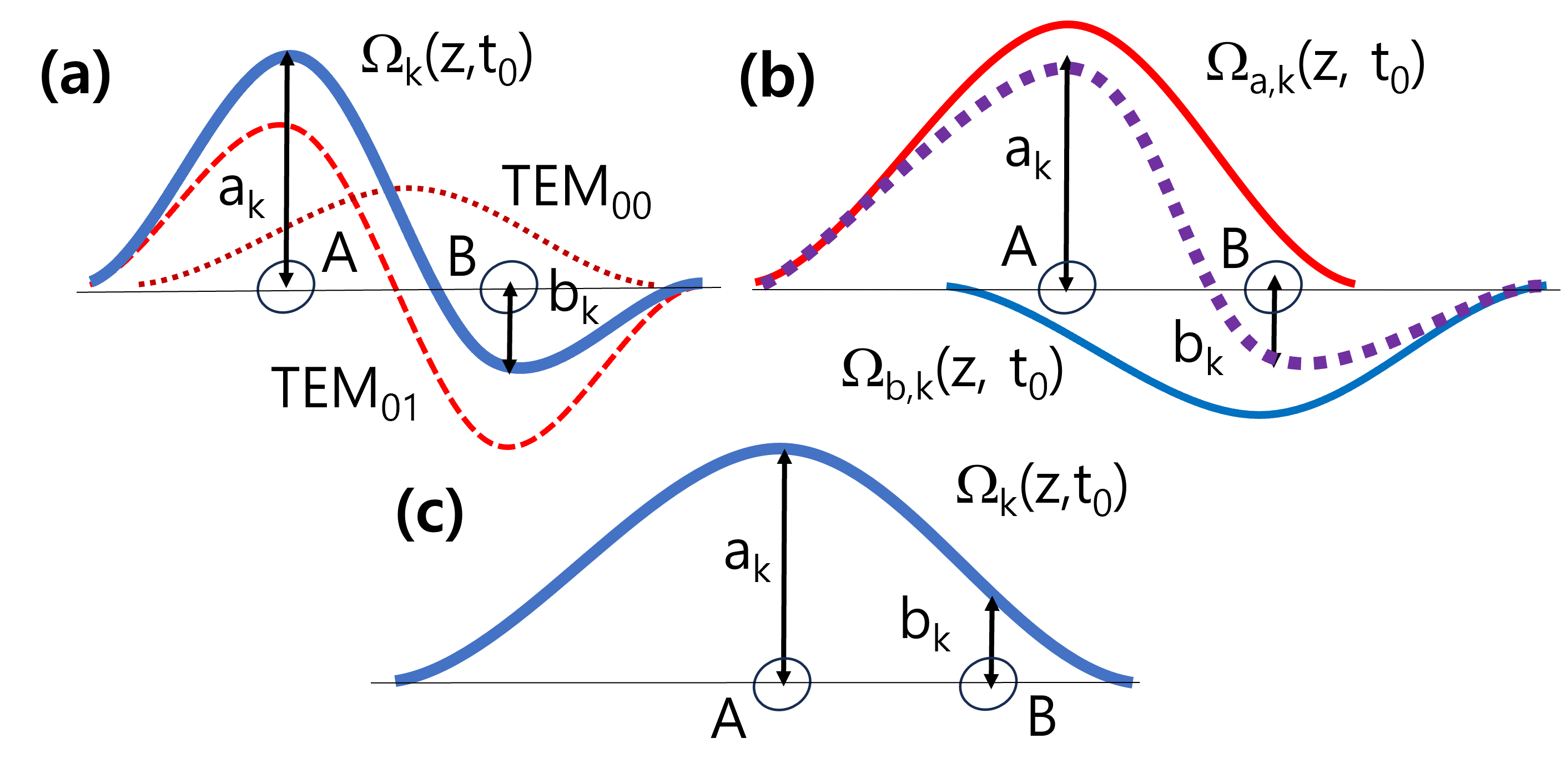}  
\caption{Diagram showing the spatial profile of the pulses at $t_0$ acting on the non-independent qubits
for different implementations of our scheme. In (a) the qubits are driven by a linear superposition
of TEM$_{00}$ andn TEM$_{01}$ modes of light, focused midway between the atoms, such that the amplitude 
of the field at qubit $A$ and $B$ is given by the desired controlled values, $a_k$ and $b_k$. 
In (b) we achieve the same level of control by acting with two Gaussian beams focused at each atom.
When the amplitudes $a_k$ and $b_k$, and hence their ratio $x_k$, can be positive as in this work,
it is possible to use a wide beam, centered on one qubit, to achieve the desired control, as
shown in (c).}
\label{scheme}
\end{figure}

The spatial control is encoded in ${\bf e}_k$ and can be achieved by different means. In \cite{Sola_Nanoscale2023} we proposed the use of hybrid modes of light.
A possible generalization for spatially non-orthogonal pulses in any
configuration, may require more complex structured light \cite{Murty_AppOpt1964,Forbes_NPhoto2021,Rubinsztein_Dunlop_JOpt2016}, 
such as those sketched in Fig.\ref{scheme} (second row).
A simpler laboratory implementation, shown in the third row, can be achieved using
a superposition of overlapping phase-locked Gaussian modes~\cite{KumarReddy_Optica2022,Chen_APE2012} centered at different qubits ,
instead of a single field, for each pulse in the sequence.
%
%
In the simplest setup, we will consider just two qubits separated at a distance $R$,
shone by two lasers at each step $k$ of the sequence, 
${\Omega}_{ak}(\vec{r},t)$ and ${\Omega}_{bk}(\vec{r},t)$, 
each focused on a qubit, but with waistbeams that span both.

We want the lasers to act with spatial coefficient $a_k$ at qubit $a$ and $b_k$
at qubit $b$.
If the beams are Gaussian (but any form is valid), 
and both lasers have the same time-dependence
given by the function of time $f(t)$, the sum of both gives the local field
at its peak,
${\Omega}_k(\vec{r}_a,t_{0k}) = 
{\Omega}_{ak}(\vec{r}_a,t_{0k}) + {\Omega}_{bk}(\vec{r}_a,t_{0k})
= \left[ \widetilde{\Omega}_{ak} + \theta \widetilde{\Omega}_{bk}\right]
= a_k\widetilde{\Omega}_{0k}$, where Rabi frequencies with
tilde represent their values at peak amplitude,
and $\theta = e^{-\alpha R^2}$ ($\alpha$ measures the beam's waist).
Here we have assumed that the spatial profile of the lasers is the same 
for all the pulses in the sequence, as will be the case in most laboratory implementations. Correspondingly, 
${\Omega}_k(\vec{r}_b,t_{0k}) = 
{\Omega}_{ak}(\vec{r}_b,t_{0k}) + {\Omega}_{bk}(\vec{r}_b,t_{0k}) =
\left[ \theta \widetilde{\Omega}_{ak} + \widetilde{\Omega}_{bk}\right]
= b_k\widetilde{\Omega}_{0k}$.

The geometrical factors can be arranged as a column (row) vector $\vec{e}_k$ 
with components $a_k$, $b_k$. In addition, we can define the column vector of field components 
$\vec{\cal E}_k = \left( \widetilde{\Omega}_{ak}, \widetilde{\Omega}_{bk} \right)$, 
and the spatial overlap matrix 
\begin{equation}
    {\sf S} = \left( \begin{array}{cc} 1 & \theta \\ \theta & 1 \end{array} \right)
\end{equation}
such that 
$\widetilde{\Omega}_{0k} \vec{e}_k = {\sf S} \vec{\cal E}_k$ and
$\vec{\cal E}_k = \widetilde{\Omega}_{0k} {\sf S}^{-1} \vec{e}_k$,
\begin{equation}
\left( \begin{array}{c} \widetilde{\Omega}_{ak} \\ \widetilde{\Omega}_{bk} \end{array} \right)
    = \frac{\widetilde{\Omega}_{0k} }{1 - \theta^2}  
    \left( \begin{array}{cc} 1 & -\theta \\ -\theta & 1 \end{array} \right)
    \left( \begin{array}{c} a_k \\ b_k \end{array} \right)
\end{equation}
which gives
\begin{equation}
\frac{\widetilde{\Omega}_{bk}}{\widetilde{\Omega}_{ak}} = \frac{x_k - \theta}{1 - \theta x_k} 
\end{equation}
where $x_k = b_k/a_k$ is the ratio of the geometrical factors.
Whenever $\theta > x_k$, assuming $x_k \le 1$, the ratio is negative. 
This can be achieved by
controlling the relative phase between the pulses.
For $x_k \le 1$, $\left| \widetilde{\Omega}_{bk} / \widetilde{\Omega}_{ak} \right| 
< \left| b_k / a_k \right|$.
Under certain conditions, it is possible (and it might be more economic) to
use a single field with more complex spatial structure, such as structured light,
instead of a superposition of overlapping Gaussian pulses.
Finally, for two-qubit of few qubit systems, and positive
relative ratios, it is possible to perform the operation with a 
single broad pulse, controlling the relative positions of the
atoms with respect to the pulse waistbeam.
In the symmetric arrangement, the pulse should be focused at mid 
distance between the atoms.  

Using superposed Gaussian beams, it is always possible to extend this procedure to more than $2$ qubits,
controlling the geometrical factors by controlling the ratio of the
peak amplitudes of the fields (as well as the pulse phases).
In the general case, one needs to define a different $\theta_{ab}$ for
each pair of qubits.
The matrix ${\sf S}$ is always invertible, as long as $\theta_{ab} \ne 1$, which 
would imply that two qubits occupy the same space.

In fact, one can use the superposition of Gaussian pulses as a technique to
remove the effect of one pulse over an unwanted qubit, if we
want to work with independent qubits even when $\alpha\sim\!R^{-2}$.
In this case, the goal is to make $x_k = 0$, for which 
$\widetilde{\Omega}_{bk} = -\theta \widetilde{\Omega}_{ak}$ 
and hence ${\Omega}_k(\vec{r}_b,t_{0k}) = 0$.

\section{Single pulse protocols}

\begin{figure}
\includegraphics[width=9cm]{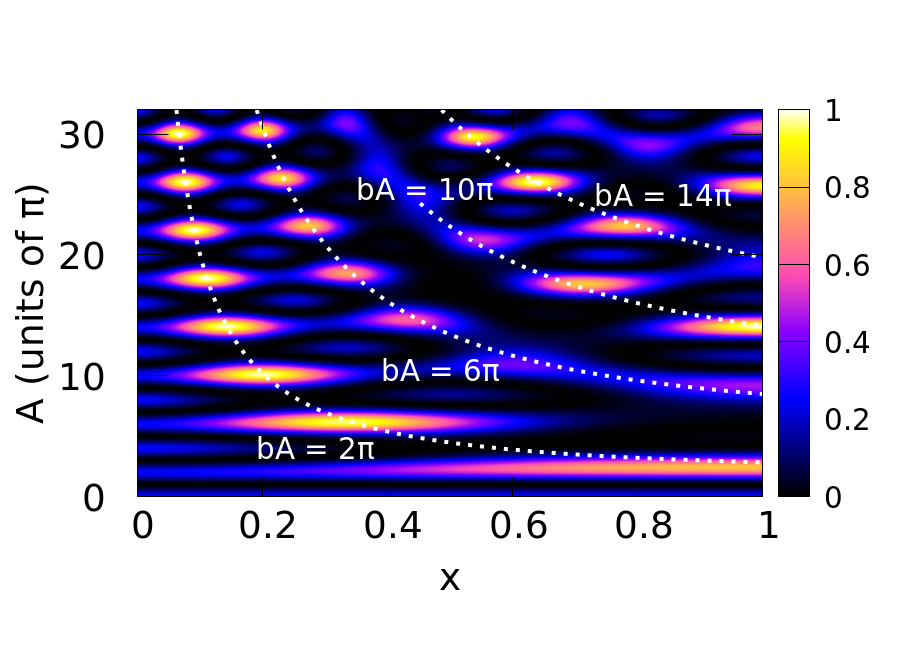}
\caption{Map of the fidelity for the CZ gate as a function of the pulse area and the ratio between the
geometrical factors, for protocols based on a single pulse acting simultaneously on both
qubits. In dashed lines, we show the protocols for which the action of the laser is
minimal in the qubit $b$. The peaks appear at approximate solutions of a Diophantine equation.}
\label{Dioph1}
\end{figure}

One of the advantages of working with non-independent qubits is that it is possible to use shorter 
pulse sequences. In principle, there are
enough control knobs to implement an entangling gate with a single pulse sequence.

For the CZ gate, we use the unconventional (but equivalent) gate definition,
where the amplitudes in each computational state, except the $|11\rangle$, 
experience a $\pi$ shift at the end of the gate.
We calculate the fidelity as
\begin{equation}
    F = \frac{1}{16} \left( -U^A_{11} -U^B_{11} -U^V_{11} + 1 \right)^2
\end{equation}
where every term $U^S_{11}$ is the first matrix element of Eq.(\ref{UV}).
For every subsystem, $S$ of states coupled by the radiation, starting 
from the different computational states, one 
must then achieve $\cos\left( \theta^S\right) = -1$.
These probability amplitudes 
correspond to so-called $0$-loop processes \cite{previouswork}, 
where the
amplitude stays solely on the computational basis by 
the end of the pulse.
For a single pulse dynamics, only $0$-loops can realize the gate.
However, it is very simple to prove that
$0$-loops can never be exactly achieved for the $3$ subsystems with a 
single pulse, so the gate mechanism cannot yield perfect fidelities even in the absence of noise or perturbations. 
The proof is simple to sketch.
\begin{proof}    
Let the system have $2$ qubits. 
For perfect fidelity, the following conditions must be satisfied:
\begin{eqnarray}
\cos (A/2) = -1 \rightarrow \sqrt{a^2+b^2}A =
(4l+2)\pi, l \in \mathbb{Z} \nonumber \\
 \cos (a A/2) = -1 
\rightarrow aA = 
(4l^\prime+2)\pi, l^\prime \in \mathbb{Z} \nonumber \\
 \cos (b A/2) = -1 
\rightarrow bA = 
(4l^{\prime\prime}+2)\pi, l^{\prime\prime} 
\in \mathbb{Z} 
\label{diopheq}
\end{eqnarray}
where we used normalized structural factors.
%
It is not possible to fulfill all the required conditions of Eq.(\ref{diopheq}) at the same time:
Calling $p=4l+2$, $n=4l^\prime+2$, $m = 4l^{\prime\prime}+2$, 
squaring the argument of the
third condition, and comparing with the first two conditions, we 
obtain the relation between the integers $m, n, p$: $m^2 + n^2 = p^2$.
Equations like this that require integer solutions are generically called {\em Diophantine equations}.
They have an infinite number of solutions. However, it can be easily shown that the solutions
cannot be constrained such that all $m, n, p$ are of the form $2, 6, 10, \ldots 4l+2$.
For, let $p > n \ge m$, $m^2 = p^2 - n^2 = (p+n)(p-n) = 16(l+l^\prime+1)
(l-l^\prime)$, while by directly squaring, 
$m^2 = 16\left( l^{\prime\prime}\right)^2 + 16 l^{\prime\prime} + 4$.
Dividing both sides by $16$ we have
$\left(l^{\prime\prime}\right)^2 + l^{\prime\prime} + 0.25 =
(l+l^\prime+1)(l-l^\prime)$.
The left-hand side cannot be integer, while the right-hand side is always integer. 
\end{proof}

It can be shown that the same restrictions apply to all $2$-qubit entangling gates.
This issue becomes more pronounced as the number of qubits increases.
For instance, with $3$ qubits, we have $3$ $V$ subsystems and $3$ two-level systems where the previous Diophantine approximate solutions must hold,
in addition to a tripod system, 
which adds another equation like $m^2 + n^2 + p^2 = q^2$, that does not hold solutions for $m, n, p, q$ integers of the type $2, 6, \ldots, 4l+2$ or similar.

However, while it is not possible to achieve perfect fidelity, 
the Equations (\ref{diopheq}) can be in principle fulfilled up to
any desired accuracy. 
For instance, in the CZ gate, 
$14^2 \approx 10^2 + 10^2$ with a relative error of
approximately $4/200 \approx 2$\%, so that an approximate solution exists using 
equal structural factors in the qubits ($a = b = 1/\sqrt{2}$) and a pulse area of $A \sim 14\pi$, which leads to a fidelity $F = 0.992$.
In Fig.\ref{Dioph1} we show a map of the fidelity of the gate as a function of the pulse
area $A$ and the ratio of the geometrical factors, $x = b/a$.
Because the role of the geometrical factors is equivalent (the fidelity is the same
for $x$ and $x^{-1}$), we only show the map for $x \le 1$.
The density of high-fidelity protocols increases for small $x$ (alternatively, $x \gg 1$).
The simplest solutions involve $bA = 2\pi$. 
For large $A$ and small $b$, $\sqrt{1-b^2} \approx 1$ and $aA \approx A$. 
This gives the series of solutions shown by the white dotted line
in Fig.\ref{Dioph1}, where
$bA = 2\pi$, from which
\begin{equation}
    bA = \frac{x}{\sqrt{1 + x^2}}A = 2\pi \longrightarrow
    A = 2\pi \frac{\sqrt{1 + x^2}}{x}
    \label{maxA}
\end{equation}
A similar equation must be satisfied by $aA$.
Dividing both, we obtain the values of $x$ at which the fidelity
is maximized,
\begin{equation}
     x_\mathrm{op} = \frac{b}{a} = \frac{bA}{aA} = 
     \frac{4l^{\prime\prime} + 2}{4l^\prime + 2} 
    \label{xopt}
\end{equation}
For the smallest possible local area in qubit $b$, $bA = 2\pi$ 
($l^{\prime\prime} = 0$),
$x_\mathrm{op}$ lie in the sequence of inverse odd numbers,
$x_\mathrm{op} = 1 / (2l^\prime +1)$.
To fully optimize the gate, the contribution of the $3$ terms
$U^A_{11}, U^B_{11}, U^V_{11}$ must be maximized, for which
the optimal pulse area must be slightly corrected as
the average between the value expected from 
Eq.(\ref{maxA}) with $x_\mathrm{op}$, 
and the value of the area that maximizes the $U^V_{11}$ term,
\begin{equation}
    A_\mathrm{op} = \left( 2l + 1 + \sqrt{\left( 2l^\prime + 1\right)^2 
    + \left(2l^{\prime\prime} + 1\right)^2} \right) \pi
    \label{Aopt}
\end{equation}
where $l \geq l^\prime \geq l^{\prime\prime} \in \mathbb{Z}$.
The protocol with smallest possible area ($l,l^\prime,l^{\prime\prime} 
= 0,0,0$)
is achieved with $A_\mathrm{op} = 2.4\pi$ at $x_\mathrm{op} = 1$ 
giving a relatively low fidelity of $F = 0.804$. 
The second maxima, at $A = 6.17\pi$ with $x = 1/3$, gives already a fidelity
$F = 0.968$.
For very large integers, the relative error can be as small as desired by increasing the pulse area, properly adjusting the ratio of the geometrical factors following Eq.(\ref{xopt}) and the area with Eq.(\ref{Aopt}).
Some results are obtained in Fig.\ref{Fnoise}.
However, as discussed in Sec.V, taking into account the effect of
fluctuations in the parameters due to shot-to-shot noise, can
shift the maximum fidelities to the lower pulse area protocols.

\section{Two-pulse protocols}

\begin{figure}
\includegraphics[width=9cm]{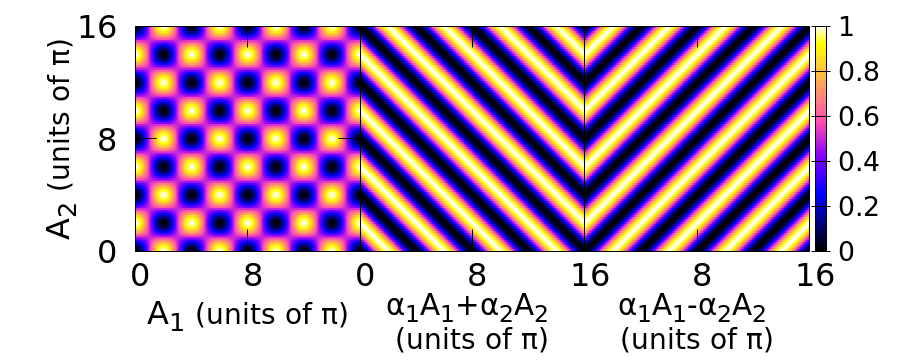}
\caption{The fidelity for protocols based on two-pulse sequences, as a function of the
pulse areas, inherit the properties of $-\cos(\theta_1)-\cos(\theta_2)$ (left) and 
$-\cos(\theta_1 \pm \theta_2)$ (center and right). 
We represent the cosine scaled and
shifted as $(-\cos x+1)/2$ so that its range is between $0$ and $1$, like the fidelity.}
\label{patterns}
\end{figure}

For two-pulse sequences, the time-evolution operator for the $A$ and
$B$ subsystems has two terms
\begin{equation}
 U^{S^\prime}_{11} = \cos\left(\alpha_2 A_2/2\right) \cos\left(\alpha_1 A_1/2\right)
- \sin\left(\alpha_2 A_2/2\right) \sin\left(\alpha_1 A_1/2\right)
\label{U2p}
\end{equation}
where $S^\prime = A,B$, $\alpha = a,b$, 
and the subscript refers to the pulse order. 
The first term is responsible for a gate mechanism based on a
$0$-loop, as in single-pulse sequences. The second-term accounts
for another mechanism that prepares the gate, the so-called one-loop,
where the first
pulse excites the population to the Rydberg state and the second
pulse takes the population back to the computational basis.
For this to happen, the GPA must be an odd multiple of $\pi$.
In the $V$ subsystem, the second term is scaled by the product
of the geometrical factors,
$U^V_{11} = \cos\left(A_2/2\right) \cos\left(A_1/2\right)
- {\bf e}_2{\bf e}_1 \sin\left(A_2/2\right) \sin\left(A_1/2\right)$
(${\bf e}_2{\bf e}_1 = a_1 a_2 + b_1 b_2$).
For each subsystem, it is in principle possible to have gate mechanisms 
that behave as 0-loops, a 1-loops, or superpositions of both.
However, because of the ${\bf e}_2{\bf e}_1$ factor, $U^V_{11}$
can only be close to $-1$ if it follows a 0-loop, unless
${\bf e}_2{\bf e}_1 = \pm 1$, that is, if the structural vectors
are aligned or anti-aligned, constraining the
parameters to be $x_2 = \pm x_1$.

\begin{figure}
\includegraphics[width=9cm]{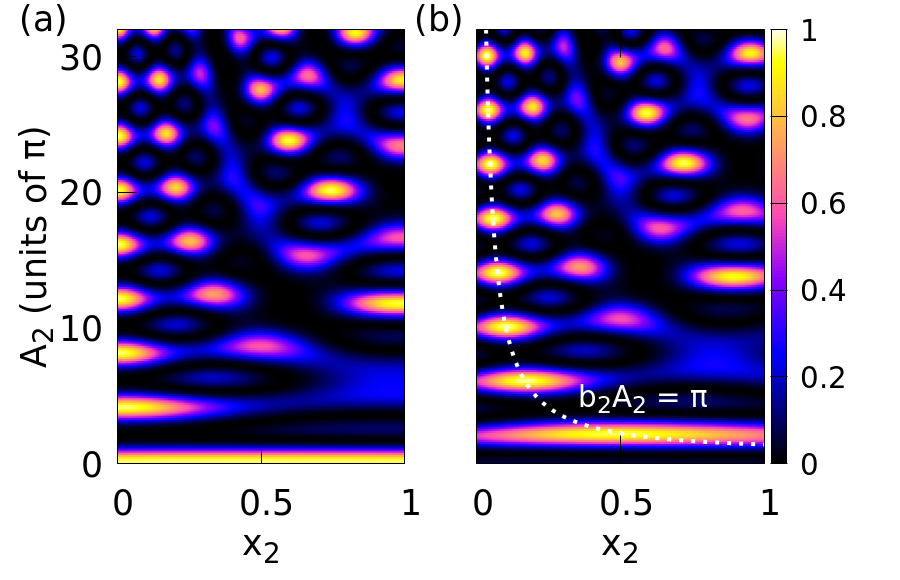}
\caption{Fidelity of the gate for two-pulse protocols as a function of $x_2$ and $A_2$.
In (a) we choose $A_1 = 6\pi$ and $x_1 = 1/3$, which are parameters that prepare a high-fidelity 
gate in the absence of the second pulse, based on a 0-loop mechanism for all subsystems.
In (b), $A_1 = 4\pi$ and $x_1 = 1/4$, so that the gate follows a 1-loop mechanism
for the B subsystem.
The maps are very similar to those of 
single-pulse sequences but with displaced areas and
ratios of the geometrical factors.}
\label{dioph-x2A2}
\end{figure}

We will first analyze 0-loop protocols, which are a natural
extension of single-pulse-based mechanisms.
For 0-loop protocols in the $V$ subsystem, 
$\cos\left(A_2/2\right) \cos\left(A_1/2\right) = -1$, which
force 
$A_1 = (4l+2)\pi$ and $A_2 = 4m\pi$
($l, m \in \mathbb{Z}$) or vice versa,
forming the checkered pattern 
of the map of protocols as a function of the pulse areas [see Fig.\ref{patterns}],
which was found in Sola et al.\cite{previouswork}
using optimization algorithms.

In Fig.\ref{dioph-x2A2}(a) we show the fidelity map as a
function of $A_2$ and $x_2$, after choosing $A_1 = 6\pi$
and $x_1 = 1/3$, which are valid parameters in a single-pulse protocol.
Hence, $A_2 = 0$ is always a possible solution.
In addition, all areas of the form $A_2 = 4m$ ($m \in \mathbb{Z}$)
provide high-fidelity gates. As the choice of $x_1$ forces the A and B subsystems
to follow a 0-loop mechanism (since $\cos\theta^{S^\prime}_1 = -1$), 
then $x_\mathrm{op} = 4m^{\prime\prime}/4m^\prime$. 
Obvious solutions of the corresponding Diophantine equations
show up at every $m^\prime$ for $m^{\prime\prime} = 0$ 
(since then $m^2 = (m^\prime)^2$ exactly), but also,
{\em e.g.} at $m = 5, m^\prime = 4, m^{\prime\prime} = 3$,
for which $x_{op} = 0.75$, etc.

In Fig.\ref{dioph-x2A2}(b) we choose $A_1 = 4\pi$ and 
$x_1 = 1/4$. Solutions exist for all areas of the second pulse
of the form $A_2 = (4m+2)\pi$. Now $b_1 A_1 \approx \pi$,
$a_1 A_1 \approx 4\pi$, so the first pulse opens a 1-loop
mechanism for the B subsystem, and a 0-loop mechanism for the
A subsystem. Then the sequence of fidelity peaks must occur
at $x_{op} = (2m^{\prime\prime}+1)/(4m^\prime + 2)$ 
(for all $m^\prime, m^{\prime\prime} \in \mathbb{Z}$.
For the smallest possible $m^{\prime\prime} = 0$, 
$b_2A_2 = x_2 A_2/ \sqrt{1 + x_2^2} \approx \pi$ and
hence $A_2 = \pi\sqrt{1 + x_2^2}/x_2$.
This is the dotted line shown in Fig.\ref{dioph-x2A2}(b)
for which high-fidelity peaks show up at
$x_{op} = 1/2, 1/6, 1/10, \ldots, 1/(4m^\prime+2)$.

It is important to note, however, that any superposition of
mechanisms can occur in the A and B subsystems.
As long as the pulse areas $A_1$ and $A_2$ alternate as
$(4l+2)\pi$ and $4m\pi$ or vice versa, it is always
possible to find high-fidelity protocols for any $x_1$, 
because from Eq.(\ref{U2p}),
$U^S_{11} = \cos\left(\theta_1 \pm \theta_2\right)$ ($\alpha = A,B$), 
with $\theta_k = \alpha_k A_1/2$. The minus sign inside the
cosine applies when
the ratios ($x_1$ and $x_2$) or areas ($A_1$ and $A_2$) change signs.
There will be always values of $x_1$, $x_2$ 
(or more precisely, of $a_1 A_1 + a_2 A_2$ and $b_1 A_1 + b_2 A_2$), for which
$U^S_{11} = -1$ for the choice of pulse areas $A_1, A_2$ that make 
$U^V_{11} = -1$. Depending on $x_1$ and $x_2$, the A and B
subsystems belong to a continuous range of mechanisms, from 0-loops
to 1-loops, passing through any combination.

\begin{figure}
\includegraphics[width=9cm]{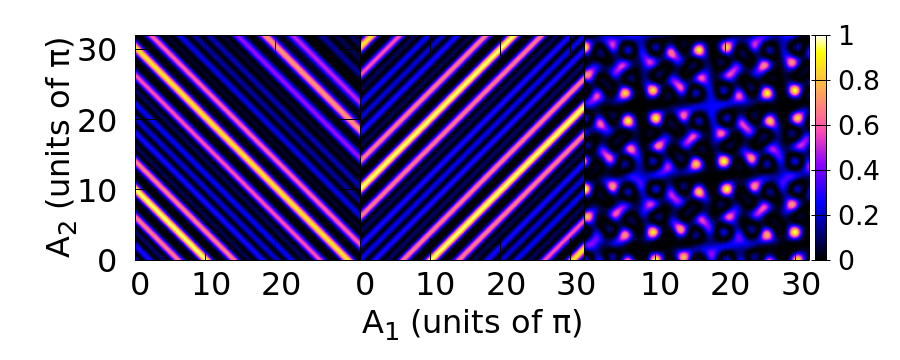}
\caption{Fidelity map as a function of the pulse areas $A_1$ and $A_2$ for 
(a) aligned structural vectors (with $x_1 = x_2 = 1/5$), (b) anti-aligned
structural vectors ($x_1 = -x_2 = 1/5$), and (c) orthogonal structural
vectors (with $x_1 = -1/x_2 = 1/5$).}
\label{dioph-A1A2}
\end{figure}


Can this realization of every possible mechanism include
the V subsystem? Indeed, if the structural vectors are
aligned or anti-aligned, ${\bf e}_1 = \pm{\bf e}_2$, for
which $x_2 = \pm x_1$, then the three terms $U^S_{11}$
($S = A,B,V$) behave as Eq.(\ref{U2p}), which can be
written as $\cos\left( \theta^S \right)$,
with $\theta^V = (A_1 \pm A_2)/2$, 
$\theta^{S^\prime} = (\alpha_1 A_1 + \alpha_2 A_2)/2$.
These are exactly the same equations as in the single-pulse
sequence, except that now the argument depends on the sum
of pulse areas,
\begin{equation}
    A_T = A_1 \pm A_2 = (4n+2)\pi, \,\,\, n \in \mathbb{Z}
\end{equation}
where the plus sign applies for aligned vectors and the minus,
for anti-aligned vectors.
So every combination of pulse areas that sums $(4n+2)\pi$
can generate a high-fidelity gate, where the mechanism can
be any superposition of 0-loops and 1-loops for all the different
subsystems.

\begin{figure}
\includegraphics[width=9cm]{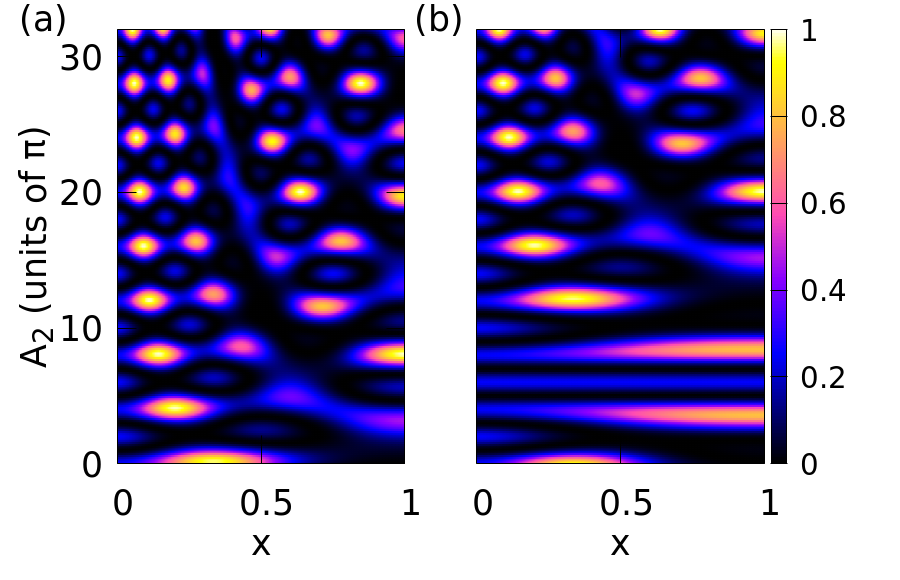}
\caption{Fidelity map as a function of the second pulse area $A_2$ and the
ratio of geometrical factors $x$ for aligned (a) $x_2 = x_1$, and 
(b) anti-aligned ($x_2 = -x_1$) structural vectors. For the figure, we
choose $A_1 = 7\pi$.}
\label{dioph-xA2}
\end{figure}

In Fig.\ref{dioph-A1A2} we show the fidelity map as a
function of the pulse areas $A_1$ and $A_2$ for 
$x = x_1 = x_2 = 1/5$ (left) and $x = x_1 = -x_2 = 1/5$ (center).
There are high-fidelity straps for pulse areas
that sum $(4n+2)\pi$, but not for all values of $n$.
The actual maximum fidelity observed and its location
depends on the choice of $x$.
The direction of the straps depends on whether the
vectors are aligned or anti-aligned.
These patterns inherit the properties of $-\cos(\theta_1\pm\theta_2)$
shown in Fig.\ref{patterns}.
In Fig.\ref{dioph-xA2} we show the fidelity map as a
function of $x$ and $A_2$, where we fixed $A_1 = 7\pi$, 
for both aligned (a) and anti-aligned (b) vectors.
As observed, $A_2 = (4n-5)\pi$. The solution that
appears at $x = 0.2$ corresponds to $A_2 = 3\pi$ (the sum of
areas equals $10\pi$). Allowing $x$ to change, one can
typically find high-fidelity protocols for any possible
valid $n$, and hence for any $A_2$.\footnote{
In spite of the fact that one can find protocols that
maximize the fidelity for any type of mechanism,
using optimal control algorithms in ref~\cite{previouswork}
we only found protocols for which the $V$ subsystem is a 0-loop.
An intriguing question is why is this so. We believe that the reason
is rooted in the peak fidelities that can be achieved by the different
mechanisms under moderate pulse areas.
Because the fidelity of 
protocols based on aligned structural vectors behaves as in
single-pulse protocols, one needs $A_T \ge 14\pi$ (for $l^\prime = l$,
$l^{\prime\prime} = 0$) to achieve $F \ge 0.99$, while
$A_T \ge 38\pi$ (for $l^\prime = l$,
$l^{\prime\prime} = 0$) to achieve $F \ge 0.999$, which was
the threshold chosen in~\cite{previouswork} to analyze the chosen
optimal protocols. Numerically, one finds many
protocols with higher fidelity and lower pulse areas when the
pulse areas are correlated as $A_1 = (4l+2)\pi$ and $A_2 = 4m\pi$ 
or vice versa, imposing a 0-loop in the $V$ subsystem. Only
when one analyzes the set of optimal protocols with lower fidelities
one observes protocols with mechanisms that do not use the 0-loop
for the $V$ subsystem.}


Finally, it is even possible to find optimal protocols where
the structural vectors are orthogonal, ${\bf e}_1 {\bf e}_2 = 0$.
They imply a superposition of the aligned and anti-aligned
vectors, for which the fidelity map looks like the pattern
observed in Fig.\ref{dioph-A1A2}(right).
The fidelity peaks form now a rotated lattice. 
The peaks are a distance of $4\pi$ apart, and the angle of
the lattice depends on the choice of $x$.
These are the solutions explored in the so-called 
SOP (symmetrical orthogonal protocol), shown in 
reference~\cite{Sola_Nanoscale2023}.

\section{Evaluating the effects of noise}

To analyze in detail all the effects of noise on the proposed schemes, one needs
to better define the setup of the system, choosing very concrete parameters
for the lasers and atomic traps, which is outside the scope of this work.
%
Our analytical approach follows from an approximate Hamiltonian from which
we can obtain the time-evolution operator, so we cannot incorporate the sources
of noise at the level of the dynamical description.
From the physical point of view, the schemes shown here operate using the
Rydberg blockade, so one can expect a similar sensitivity to the fluctuation
of the laser frequency, the spontaneous emission, and the thermal motion of the atoms,
as reported elsewhere. \cite{Ahn_PRA2019} 
However, because the atoms are much closer, the dipole blockade is much larger
and the 
pulses much shorter (operating, in principle in tens of nanoseconds) and
much more intense, the phase-induced detunings or changes in population due to spontaneous decays, which are the main sources of errors in microsecond
experiments, become almost negligible in our setup.
Mainly shot-to-shot fluctuations, rather than decoherence, will have some impact on the fidelities.

Herein, we develop a simple model to evaluate the impact of fluctuations in 
the pulse energy (hence pulse areas) and geometrical factors on the fidelity for the CZ gate in two-qubit systems, using two partially overlapping pulse beams centered at each qubit.

The impact of amplitude fluctuations over the pulse areas is direct.
For a pulse with intensity $I_0 = c \epsilon_0^2$, given that the area is $A_0 = \mu \epsilon_0 S_0/\hbar$, where $S_0$ is a shape factor, 
neglecting fluctuations in the pulse duration (or rather, subsuming the effect on the
peak intensity fluctuation),
the relative error in the pulse areas is
\begin{equation}
    \delta A_0 \equiv \Delta A_0 / A_0 = \Delta I_0 / 2I_0
    \label{noise1}
\end{equation}
Using stabilized microsecond pulses, $\delta I_0$ can be estimated as $\sim\!3$\% or smaller.

Fluctuations in the geometrical factors depend both on fluctuations in the laser
amplitudes as well as on the thermal motion of the atoms.
For the parameter $b_k$ obtained by a superposition of beams
\begin{equation}
  b_k = \frac{\epsilon_{bk}}{\Omega_{0k}} + \theta \frac{\epsilon_{ak}}{\Omega_{0k}}  
    \label{bsup}
\end{equation}
where $\Omega_{0k} = \sqrt{\epsilon_{bk}^2 + \epsilon_{ak}^2}$, we separate
the dependence on $\epsilon$ from the dependence on $R$ through 
$\theta = \exp(-\alpha R^2)$, as $\Delta b_k^2 = (\Delta b_k^\prime)^2 + (\Delta b_k^{\prime\prime})^2$,
where
$$ (\Delta b_k^\prime)^2 = \left( \frac{\partial b_k}{\partial \epsilon_{ak}} \right)^2
(\Delta \epsilon_{ak})^2 + \left( \frac{\partial b_k}{\partial \epsilon_{bk}} \right)^2 
(\Delta \epsilon_{bk})^2  \ .$$
Assuming that the relative errors in the fields are similar, 
$\delta \epsilon_{ak} = \delta \epsilon_{bk} = \delta I_0 / 2$, 
\begin{eqnarray}
(\Delta b_k^\prime)^2 & = &
\frac{1}{4}\left\{ \frac{\epsilon_{ak}^2\theta^2 + \epsilon_{bk}^2}{\Omega_{0k}^2}
+ \left( \frac{b_k}{\Omega_{0k}} \right)^2 \frac{\epsilon_{ak}^4 + \epsilon_{bk}^4}{\Omega_{0k}^2} \right\} (\delta I_0)^2 \nonumber \\
 & \le & \frac{1}{2}b_k^2 (\delta I_0)^2 \nonumber
\end{eqnarray}
On the other hand,
$$(\Delta b_k^{\prime\prime})^2 \!= \!\left( \frac{\partial b_k}{\partial \theta} \right)^2
\!\!\left( \frac{d \theta}{d R} \right)^2 \!\! (\Delta R)^2 = 
\left( \frac{\epsilon_{ak}}{\Omega_{0k}} \right)^2 \!\!\left( 2\alpha R \theta \right)^2 \!(\Delta R)^2 \ .$$
Since $2\alpha R^2 \theta\sim 1$, $(\Delta b_k^{\prime\prime})^2 \sim \theta^2 (\delta R)^2$,
from which
\begin{equation}
    \Delta b_k^2 \sim  \frac{1}{2}b_k^2 (\delta I_0)^2 + \theta^2 (\delta R)^2
\end{equation}
Equally, for the $a_k$ term, we have
\begin{equation}
    \Delta a_k^2 \sim  \frac{1}{2}a_k^2 (\delta I_0)^2 + x^2 \theta^2 (\delta R)^2
\end{equation}
so
\begin{equation}
\delta x^2_k = \delta b_k^2 + \delta a_k^2 = (\delta I_0)^2 + \left(2 + \frac{1}{x_k^2 (x_k^2 +1)}
\right) \theta^2  (\delta R)^2
\label{noise2}
\end{equation}
where we observe that $\Delta x_k$ depends on $x_k^{-1}$ (because $\delta x_k$ depends on $x_k^{-2}$),
so we expect the error to be larger for protocols that work with small $x_k$. This is why the unwanted
presence of a second qubit can damage the fidelity of a scheme based on independent qubits. These 
detrimental effects can be somehow reduced in the SOP. 

To evaluate the error in $\delta R$, 
we use a simple estimation assuming a diffusion model for the dispersion of the atoms,
$ \Delta R \sim \sqrt{2D t_g}$,
where $t_g$ is the gate duration and $D$ the diffusion coefficient.
In~\cite{Ahn_PRA2019}, 
working with atoms separated $5\mu$m and using gates that operate in $\sim\!5\mu$s
at $\sim\!25~\mu$K, the authors evaluate $\Delta R$ as $\sim\!50\,$nm.
If we assume that our gates operate under similar conditions (e.g. temperature) but $25$ times faster, 
that would imply $\Delta R\sim\!10$nm, 
for a relative error of $\delta R \sim\!1$\% when the atoms are approximately $1\mu$m apart, although our approximations may underestimate the error
during the measuring of the gate's state. To evaluate the effect of the
temperature, we will assume a linear dependence with the mean square
displacement, as in Brownian motion, or for classical and quantum
oscillators under certain limits~\cite{Marquardt_MP2021}. 

\begin{figure}
\includegraphics[width=9cm]{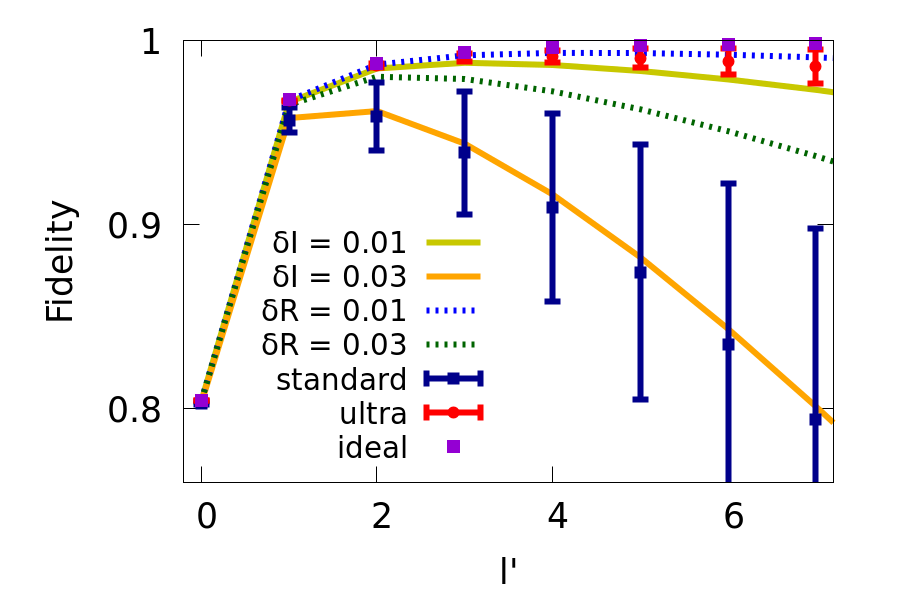}
\caption{Fidelity for single-pulse protocols with different $l = l^\prime$ 
and $l^{\prime\prime} = 0$ for different levels of noise in the parameters.
The dotted lines show the errors induced by thermal fluctuations in the
positions of the atoms with relative standard deviations $\delta R = 0.02, 
0.01$ (lower and higher curves). Solid lines show the errors induced
by fluctuations in the peak intensities of the pulse, with relative standard
deviations $\delta I = 0.03, 0.015$ (lower and higher curves).
The squares are the results in the absence of fluctuations, and
circles with error bars show the results for the fidelity in the presence
of both noise sources, with $\delta R = 0.02$, $\delta I = 0.03$. 
The error bars show the standard deviation in the fidelity, across 
a distribution of $1000$ samples.}
\label{Fnoise}
\end{figure}

We use Eqs.(\ref{noise1}) and (\ref{noise2}) to evaluate a distribution of
parameters $A$ and $x$ following the noise statistics. We also include
a distribution in the absolute phase of the lasers with $\Delta\phi =
0.01\pi$. Using a sample of
$1000$ different parameters, we evaluate
the average fidelity 
and the standard deviation for several single-pulse optimal protocols
(with $l^{\prime\prime} = 0$ and $l = l^\prime$) with different
noise contributions.
In Fig.\ref{Fnoise} we show the fidelities in the absence of fluctuations
(squares) and the average fidelities with $\delta I = 0.03$, 
$\delta R = 0.01$ ($T\sim\!25\mu$~K) and $\Delta\phi = 0.1\pi$, 
labelled as ``standard'', which are the errors reported in~\cite{Ahn_PRA2019}. 
The error bars show the standard 
deviation in the fidelity, which, for $l^\prime = 6$ reaches 
$\sigma = 0.17$. The results 
reveal that fidelity is
severely affected for protocols that use large $l^\prime$ (and $l$ and
hence $A$),
which correlate to protocols that operate with larger Rabi frequencies
and smaller ratios of the geometrical factors.

The effect of fluctuations in the laser amplitudes (solid lines) is quite
stronger than the effect of fluctuations on the atomic positions (dotted lines).
Although the relative error both in $A$ and $x$ is linearly proportional
to the relative error in the pulse intensities, the required precision
in the intensities should increase for protocols that use stronger fields,
as a small error in $A$ can easily shift the GPA from an odd multiple to 
an even multiple of $\pi$ (and vice versa), totally changing the excitation
mechanism. For intensity fluctuations of $\sim\!3$\%, only the lowest
area protocols ($A \le 10\pi$) survive with fidelity errors smaller than
$5$\%. It is really necessary to reduce the laser fluctuations 
to one-half of this value or lower ($1$\% in the yellow line)
to reduce the errors to less than $2$\% in protocols with
$A = 14\pi$. In Fig.\ref{Fnoise}, labelled as ``ultra'',
we also show the results using noise statistics currently available~\cite{Ahn_PRL2020}
with state-of-the-art laser stabilization ($\delta I \sim 0.007$, 
$\Delta \phi \leq 0.01\pi$) and sideband cooling ($T = 3~\mu$K),
which show 
that errors in fidelity can, in principle, be reduced to less than
$1\%$.
In fact, all the errors in such conditions
depends on $\delta I$, as practically the same results would be obtained at
$T = 30~\mu$K.


\section{Conclusions}

In this work, we have studied minimal pulse sequences that implement the CZ gate on two adjacent
and non-independent qubits with high fidelity, where the number of pulses used per qubit can be 
as small as one. 
Indeed, using structured light, in principle one can implement the gate with a single pulse.
We have proposed a possible implementation using superposed Gaussian beams, and we have
analyzed the role of parameter fluctuations induced by shot-to-shot noise.

Ultimately, the optimal parameters must be approximate solutions of Diophantine equations,
imposing strict conditions on the pulse areas and overlaps of the pulses.
While perfect fidelities can never be achieved even under ideal conditions,
the errors can be made as small as desired using intense pulses.
The use of two-pulse sequences looses the restrictions on the values of the parameters that optimize
the gate. One finds that a continuum of mechanisms, described in terms of quantum pathways,
can be used for its implementation, although strong correlations in the areas
of the pulses of the form $A_1 = (4l + 2)\pi$, $A_2 = 4m\pi$ ($l, m \in \mathbb{Z}$) or
vice versa, are typically found in optimal protocols. 

By implementing the qubits in atoms trapped at a short distance of each other (thereby boosting the dipole blockade), the goal is to speed up the gates to the nanosecond time-scale.
We found that intensity fluctuations have a much stronger impact on the fidelity
of the gates than the thermal motion of the atoms, mainly in protocols that
use large pulse areas and hence, assuming short pulses, strong fields.
Our preliminary analysis reveals that the stabilization of the lasers 
that allows to reduce the relative errors in the pulse intensities below $1\%$, may be necessary 
for the laboratory implementations of these protocols. On the other hand, 
the experiments can be performed at typical ultracold temperatures of 
$\sim 10$ $\mu$K.

While an in-depth analysis of all protocols can only be made for small pulse sequences,
we expect that the use of protocols with several pulses with similar total accumulated Rabi 
frequency, but smaller peak intensities, can result 
in higher fidelities and more robust gates.

\section*{Acknowledgements}
This research was supported by the Quantum Computing Technology Development Program (NRF-2020M3E4A1079793). IRS thanks the BK21 program (Global Visiting Fellow) for the stay during which this project started and the support from MINECO PID2021-122796NB-I00. SS acknowledges support from the Center  for Electron Transfer funded by the Korean government(MSIT)(NRF-2021R1A5A1030054)

\bibliography{minimal_sequences.bib}

\end{document}